\documentclass[prd,aps,twocolumn,superscriptaddress,unsortedaddress,amsmath,amssymb]{revtex4-1}
\usepackage{epsf}
\usepackage{graphicx}

\newcommand{\lsim}{\raisebox{-0.07cm}{$\;\stackrel{<}{{\scriptstyle \sim}}\; $} }

\begin{document}

\title{A study of the $\eta$ and $\eta^\prime$ mesons
with improved staggered fermions}


\author{Eric B. Gregory}
\affiliation{Department of Physics, University of Cyprus, P.O. Box
  20357, 1678 Nicosia, Cyprus}
\altaffiliation{Current address: 
Bergische Universit\"at Wuppertal, Gaussstr.\,20, D-42119 Wuppertal,
Germany}

\author{Alan C. Irving and Christopher M. Richards}
\affiliation{
Theoretical Physics Division, Dept. of Mathematical Sciences,
      University of Liverpool, Liverpool L69 7ZL, UK}

\author{Craig McNeile}
\affiliation{
Bergische Universit\"at Wuppertal, Gaussstr.\,20, D-42119 Wuppertal,
Germany
}

\collaboration{UKQCD Collaboration}
\noaffiliation

\begin{abstract}
We report on a high statistics lattice QCD calculation of the 
mass of the $\eta$ and $\eta^\prime$ mesons using ASQTAD improved 
staggered fermions.
The calculation used two ensembles with different lattice spacings 
and pion masses.
We also report results for $\eta$-$\eta^\prime$ mixing.
The results are in satisfactory agreement with other lattice 
calculations using other fermion formulations and with experiment,
given the the unphysical quark masses used.
We see no evidence of abnormal behavior at the lattice spacings studied.
\end{abstract}

\maketitle


\section{Introduction}
\label{se:intro}

The explanation for the large mass of the $\eta^\prime$ meson relative to the
masses of the light pseudo-scalar mesons involves
nonperturbative physics, 
such as topology and the $U(1)$ 
anomaly~\cite{Feldmann:1999uf,Shore:2007yn,Kawarabayashi:1980dp}, so these
masses present an interesting challenge for lattice QCD.
There are many phenomenological applications that
require knowledge of the properties
of the $\eta$
and $\eta^\prime$ mesons, such as their decay constants.  
For example, Di Donato et al.~\cite{DiDonato:2011kr}
have recently noted 
that a theoretical description of $\eta$-$\eta^\prime$ 
mixing is required for understanding 
the CP asymmetries
of charmed and bottom mesons with 
$\eta$ or $\eta^\prime$ in the final state.
Fleischer et al~\cite{Fleischer:2011ib} 
have recently studied the dependence
of $\eta$-$\eta^\prime$ mixing on the
decay $B^0_{s,d} \to J/\psi \eta^{(\prime)}$,
and which could be used to look for physics beyond
the standard model. 
There is currently an active experimental 
program~\cite{Moskal:2011ve}
into the properties of the $\eta$ or $\eta^\prime$ mesons
at experiments such as
WASA~\cite{Adam:2004ch}, KLOE-2~\cite{AmelinoCamelia:2010me},
and MAMI~\cite{Unverzagt:2009vm}.
Lattice QCD can in principle compute
the $\eta$-$\eta^\prime$ mixing from first principles,
so it should be able to help interpret the experiments.

The improved staggered fermion formulation, known as 
ASQTAD fermions~\cite{Orginos:1998ue,Orginos:1999cr},
has resulted in some very accurate lattice QCD calculations which
have been validated against
experiment with a high degree of 
precision~\cite{Davies:2003ik,Gregory:2009hq}. 
The results from the improved staggered fermions physics program
have been recently been reviewed by the 
MILC collaboration~\cite{Bazavov:2009bb}.

Despite the success of the phenomenology from staggered fermions,
concerns have been expressed over the theoretical basis of the formalism.
The central issue arises from the so-called \lq rooting\rq{} of the 
staggered determinant~\cite{Creutz:2007yg,Creutz:2008nk}. 
This is invoked in order to correct the number of flavors/tastes of
sea quarks which would otherwise arise.
These aspects of the staggered fermions formalism have been
reviewed by D\"{u}rr~\cite{Durr:2005ax}, 
Sharpe~\cite{Sharpe:2006re}, 
Creutz~\cite{Creutz:2007rk} and 
Kronfeld~\cite{Kronfeld:2007ek}.

The physics of the $\eta^\prime$ meson is sensitive
to the topology of QCD. Since some of the concerns about rooting
the staggered formalism are related to the 
topology~\cite{Creutz:2007rk,Creutz:2008nk,Creutz:2007rk},
the $\eta^\prime$ could be a place where incorrect
results are obtained.
The properties of the $\eta^\prime$ meson are also thought
to be 
related to the eigenvalue spectrum of the Dirac operator.
The eigenvalue spectrum of a type of improved staggered
fermions called HISQ has recently been shown to agree
with continuum 
expectations~\cite{Follana:2004sz,Follana:2005km,Donald:2011if,Durr:2004as},
which is suggestive that the physics of the $\eta^\prime$
meson is correct with improved staggered fermions.
The recent paper by Donald et al~\cite{Donald:2011if} also 
contains a rebuttal of some previously raised theoretical concerns.
A calculation of the properties of flavor singlet mesons
is a further and crucial test of the validity of the improved
staggered formalism.

The very successful physics program that uses
improved staggered fermions
has largely been independent of the properties 
of the $\eta$ and $\eta^\prime$ mesons~\cite{Bazavov:2009bb}. 
The MILC collaboration has recently used the properties
of the $\eta^\prime$ in their study 
of the topological susceptibility~\cite{Bazavov:2010xr}.
The  $\eta$ and $\eta^\prime$ mesons could indirectly
affect other lattice calculations.
For example, the $\eta$ and $\eta^\prime$ mesons are
common decay products of hadrons that decay
via the strong force, such as the 
$a_0$ meson~\cite{Aubin:2004wf,Bernard:2007qf} or
the proposed 
exotic hybrid 
mesons~\cite{Bernard:2003jd}.
However the majority of lattice QCD calculations performed
by the MILC collaboration are probably independent of the 
properties of the $\eta$ and $\eta^\prime$ mesons.

In~\cite{Gregory:2007ev} we reported on methods 
to compute disconnected
diagrams with improved staggered fermions. 
In that study we found that we needed much higher statistics
than those of a typical ensemble generated by the MILC
collaboration (at that time). This new work is based 
on generating many more configurations in order to
improve the statistics.
We have recently reported~\cite{Richards:2010ck}
results for the masses of the $2^{++}$, $0^{++}$ 
and $0^{-+}$ glueballs obtained as a byproduct of this calculation.

Lattice QCD calculations of flavor singlet quantities are 
not as well developed as those of flavor nonsinglet
quantities, because they are computationally more
expensive. There have been lattice QCD calculations of the 
$\eta^\prime$ mesons with 2+1 flavors of sea quarks, by the 
JLQCD collaboration~\cite{Aoki:2006xk}, 
RBC/UKQCD collaboration~\cite{Christ:2010dd},
Hadron Spectrum Collaboration~\cite{Dudek:2011tt},
and TWQCD and JLQCD collaborations~\cite{Kaneko:2009za}.
A preliminary study of the form factors for the semileptonic decay
of $D_s \rightarrow \eta_s$ has been carried out~\cite{Bali:2011yx}.
There is also a preliminary calculation of the masses of $\eta$ and
$\eta^\prime$ mesons, with 2+1+1 
flavors of sea quarks,
from the ETM collaboration~\cite{Ottnad:2011mp}.

There have been a few
lattice QCD calculations of the 
flavor singlet pseudoscalar meson with 
$n_f$=2 sea quarks.
However, the strange quarks in the sea play an essential role 
in lattice QCD calculations of the mass of 
the $\eta^\prime$ meson. For example 
Jansen et al.~\cite{Jansen:2008wv} found
the ground state of the flavor 
singlet pseudoscalar meson
in $n_f$ = 2 to be 0.865(65)(65) GeV compared 
with the experimental value of the $\eta$ mass
of 0.548 GeV which lattice QCD calculations with 
2+1 sea quarks should reproduce.

\section{Details of the lattice calculations}
\label{se:simulations}

\begin{table*}[tb]
\centering
\begin{tabular}{|c|c|c|c|c|c|c|c|c|c|} \hline
Name & $\beta$ & $u_0$  & $L^3 \times T$  &  $a m_l$ & $a m_s$   &
$N_{\rm cfg}^{\rm(total)}$ & $N_{\rm traj}$ & $N_{\rm cfg}^{\rm (meas)}$  & $r_0/a$ \\ 
\hline
coarse & 6.75  & 0.8675 & $24^3 \times 64$ & 0.006   & 0.03 & 5237 &
31422 &  4452 &  3.8122(74) \\ 
fine   & 7.095 & 0.8784 & $32^3 \times 64$ & 0.00775 & 0.031  & 2867 &
17202 & 2808 & 5.059(10)
\\ \hline
\end{tabular}
\caption{Summary of ensembles used in this calculation. We use the 
same convention for the quark masses as used by the MILC 
collaboration.}
\label{tb:ensemble}
\end{table*}

We used the improved staggered fermion action 
called ASQTAD~\cite{Orginos:1998ue,Orginos:1999cr}
with the tadpole improved Symanzik gauge action.
We generated two ensembles at two lattice spacings,
using  QCDOC machines~\cite{Chen:2000bu}.
The basic parameters of the lattice QCD calculations are
in Table~\ref{tb:ensemble}. We generated $N_{\rm cfg}^{\rm(total)}$
configurations, but base our final calculation on a subset of  
$N_{\rm cfg}^{\rm(meas)}$ configurations.
The RHMC algorithm was used to 
generate the configurations~\cite{Clark:2006fx,Clark:2006wp}.
The details of the tuning of the 
RHMC algorithm are in~\cite{Richards:2009thesis}.

The use of new measurement techniques
has improved the precision of many lattice QCD measurements.
However, for many applications of lattice QCD 
the use of higher statistics
is still an essential requirement. 
In particular, this is so for glueball studies~\cite{Gregory:2005yr}, 
or for anything which includes the calculation of disconnected diagrams.

Our aim was to generate 30,000 trajectories. For comparison
the ETM collaboration has typically used ensemble sizes of 5000 
or 10000 trajectories~\cite{Jansen:2008wv}. 
The length of simulation required also depends on the
autocorrelation time.
There have been some recent high statistics lattice QCD 
calculations using an anisotropic lattice~\cite{Beane:2009ky}.

In our methods paper~\cite{Gregory:2007ev} 
on the $\eta^\prime$ we found
that the correlators did not have a Gaussian distribution.
A similar observation has been made by the ETM 
collaboration~\cite{Jansen:2008wv} in their calculation
with $n_f =2 $ flavors of sea quarks. In this calculation
we again observe a 
long tail in the distribution of the disconnected
correlators. 
 \begin{figure}
\begin{center}
\includegraphics[width=83mm]{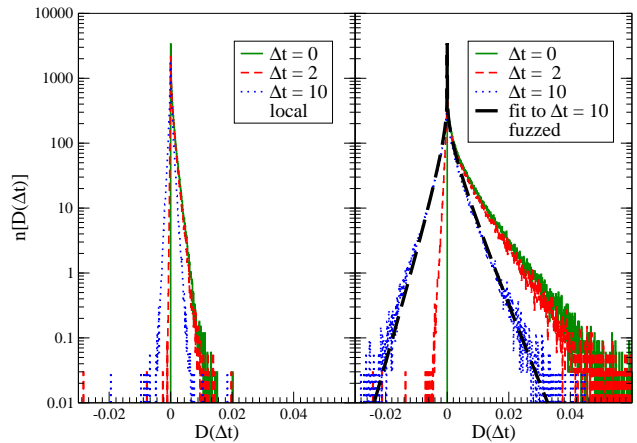}
\end{center}
 \caption {
Normalized histograms of raw disconnected correlator measurements with
local (left) and fuzzed (right) pseudoscalar source operators. Measurements
are for the light-light correlator on the coarse ensemble. Note that symmetry
increases with $\Delta t$ and the most likely value remains at 
$D(\Delta t) = 0$ for all $\Delta t$. Histograms contain 335168 measurements 
(5237 configs $\times$ 64 timeslices. The heavy dashed green curve is a 
fit of Eq.~(\ref{MBF2_eq}) to the fuzzed $\Delta t = 10$ histogram.).
 }
\label{fig:dcorr_histo}
 \end{figure}

In ~\cite{Gregory:2007ev} we pointed out that the long-tailed distribution is 
an inherent challenge to measuring disconnected correlators. Under the 
assumption (well-justified by numerical observation) that the pseudoscalar 
fermion loop operators fluctuate within a Gaussian distribution about zero,
 the product of two loop operators, measurements disconnected correlator
$d$ must fall in a long-tailed distribution,
specified by 
\begin{equation}
\label{MBF2_eq}
n(d) = A\exp(-B d)K_0(C \left|d\right|),
\end{equation}
where $K_0$ is a modified Bessel function of the second kind, 
and $A$, $B$ and $C$ are constants.

The distribution of disconnected correlator measurements will {\em always} have
its peak at zero, with the signal coming from the asymmetry of the long tails.
This behavior may, however, be masked by any binning of measurements.
We show some histograms of disconnected correlator measurements in 
Fig. \ref{fig:dcorr_histo}.

The UKQCD/RBC collaboration has, however, not observed
any  ``large statistical excursions'' in the distribution of
the disconnected correlators in an unquenched calculation with
2+1 flavors of sea quarks~\cite{Christ:2010dd}.

It might therefore be tempting to suggest that the long tail
in the distribution of the disconnected correlators
is caused by the lack of \lq lattice\rq{} chiral symmetry in the 
staggered and twisted mass fermion operators, compared
to the domain wall formalism. However, we explicitly checked
that we used sufficient numbers of noise sources so that our errors were
dominated by the gauge noise. 
The wall sources used by the
RBC/UKQCD collaboration~\cite{Christ:2010dd} 
can be thought of as using a single
noise source but with partitioning in time.
It might be that 
the distribution of the errors on 
disconnected correlators in the latter case are not dominated by the
gauge noise. This issue is discussed by
Wilcox~\cite{Wilcox:1999ab} and there is a recent
study by Alexandrou et al.~\cite{Alexandrou:2011ar}.

Recently Endres et al.~\cite{Endres:2011jm} have discussed
heavy-tailed correlator distributions in condensed
matter simulations and lattice QCD
calculations~\cite{Endres:2011mm},
and started to develop new
techniques to extract physics in such situations.

We note that the width of the disconnected correlator distribution 
increases with fuzzing of the 
sources. The taste-singlet pseudoscalar operator in the staggered formulation
has the quark and anti-quark separated by four links. It is possible that 
the introduction of the gauge links in the covariant shift or in the fuzzing
increases the width of the Gaussian distribution of the loop operators 
and hence the width of distribution of the disconnected correlator 
described by (\ref{MBF2_eq}). In other 
formulations, the distributions may possibly be narrower and less troublesome.

It can be difficult to make a good choice of strange 
quark mass to use in a lattice QCD calculation,
although the choice is getting easier
with increased experience of 2+1 calculations.
In the original calculations by the MILC collaboration,
the estimate of the strange quark mass turned out to be off 
by as much as 25\%~\cite{Bazavov:2009bb}. 
The MILC collaboration then corrected for the mismatch of the strange 
quark mass by using their
extensive partially quenched data sets and later unquenched runs. 
Their updated estimates for the strange quark mass (in lattice
units) were 0.035(7) and 0.0261(7) on the coarse
and fine ensembles respectively~\cite{Bazavov:2009bb}.
However we have used the new value of 
$m_s$ that MILC used on the coarse~\cite{Bernard:2007ps}, 
so as to interpolate with their
original value of the strange quark mass.
Indeed, our coarse ensemble has already been used
in an analysis of the strangeness content
of the nucleon, by the MILC collaboration~\cite{Toussaint:2009pz}.

We first discuss the interpolating operators and two-point functions
for flavor singlet pseudoscalar mesons using  Wilson
fermions with $n_f$ degenerate quarks. 
A longer discussion can be found in our original paper~\cite{Gregory:2007ev}.
Two-point functions 
\begin{equation}
G(x^\prime, x) = \langle O(x^\prime) O(x)  \rangle
\end{equation}
are constructed using the interpolating
operators
\begin{equation}
O(x) = \sum_{i=1}^{n_f} \overline{\psi}_i (x) \gamma_5 \psi_i (x)\, .
\end{equation}

Using standard Wick contractions, the two-point
function decomposes into a connected ($C$) and 
disconnected ($D$)
part
\begin{equation}
G(x^\prime, x) = n_f C(x^\prime, x) - n_f^2 D(x^\prime, x)
\label{eq:coor}
\end{equation}

In the staggered formalism the native four tastes of sea quarks are reduced 
to $n$ degenerate flavors through the $\frac{n}{4}$-rooting of the fermion matrix. 
The four valence tastes manifest themselves in valence loops, which come 
in too large by a factor of four. A connected correlator is essentially a 
single valence loop and requires normalizing by a factor of $\frac{1}{4}$
to get the single-flavor contribution. A disconnected correlator contains
two valence loops (along with an arbitrary number of sea quark loops).
Computing the single-flavor disconnected contribution therefore requires 
normalizing by two factors of $\frac{1}{4}$.
This factor is sometimes referred to as \lq valence
rooting\rq~\cite{Sharpe:2006re}. We include these factors implicitly in our 
definition of $D$ and $C$.

In the staggered formulation we use the Kluberg-Stern 
notation~\cite{KlubergStern:1983dg}.
The Goldstone pion operators are $(\gamma_5 \times \gamma_5)$,
and the flavor singlet pion operators are 
 $(\gamma_5 \times 1)$, where the first index refers to 
spin and the second index to taste.

The signal-to-noise ratio for the flavor singlet pseudoscalar
mesons rapidly falls with time, 
so it is important to use spatially smeared operators
to project onto the ground and first excited state as soon as
possible.
We used the fuzzing technique~\cite{Lacock:1994qx} with
a fuzzing length of 4 (this must be even to respect the symmetries
of the staggered action).

 \begin{figure}
\begin{center}
\includegraphics[scale=0.3,angle=270]{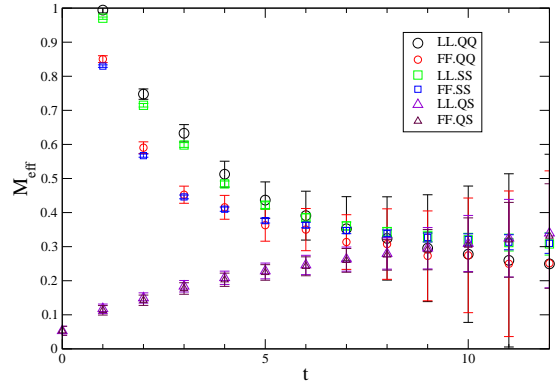}
\end{center}
 \caption {
Effective mass plot for the connected part of $(\gamma_5 \times 1)$
operators for the fine ensemble.
 }
\label{fig:mefffine}
 \end{figure}

 \begin{figure}
\begin{center}
\includegraphics[scale=0.3,angle=270]{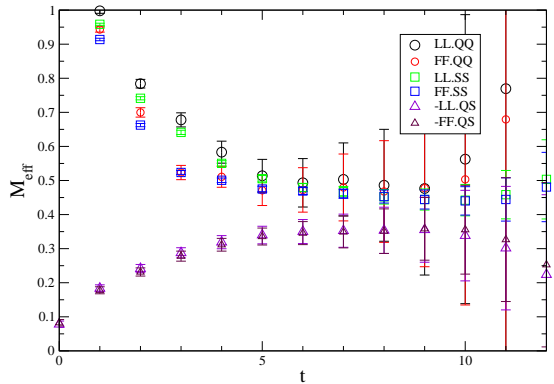}
\end{center}
 \caption {
Effective mass plot for the connected part of $(\gamma_5 \times 1)$
operators for the coarse ensemble.
 }
\label{fig:meffcoarse}
 \end{figure}

We computed a variational correlator 
matrix ${\cal C}$ using four basis states - light and strange quark pseudoscalar
correlators combined with local and fuzzed 
sinks and sources:
\begin{equation}
\label{eq:oper}
{\cal C}(t)_{ij} = \langle O_i(t) O_j(0)^\dagger \rangle\, .
\end{equation}

In Figs.~\ref{fig:mefffine} and~\ref{fig:meffcoarse}
we show the effective mass plots for the
local and fuzzed smeared operators for the 
 $(\gamma_5 \times 1)$ pion operator.
For the main $\eta$-$\eta'$ analysis, we 
use the factorizing fit model~\cite{Lepage:2001ym} 
shown in Eq.~(\ref{eq:corrmodel}),
\begin{equation}
\label{eq:corrmodel}
{\cal C}(t)_{ij} = \sum_{k=1}^{N} 
\frac{a_{ik} a_{jk} }{ 2 E_k}
( e^{- E_k t } + e^{- E_k (T - t) }     ) 
\end{equation}
where $i,j$ run from 1 to 4. Here $N$ is finite and all states are 
treated as stable. In final fits we used no Bayesian constraints.

Physics can also be extracted from the measured variational
matrix using the generalized eigenvalue 
method~\cite{Michael:1985ne,Luscher:1990ck,Blossier:2009kd}.
\begin{equation}
{\cal C}(t) v_n(t,t_0) = \lambda_n(t,t_0){\cal C}(t_0) v_n(t,t_0)
\end{equation}
where $v_n(t,t_0)$ are the generalized eigenvectors.
Typically a small time $t_0$ is chosen.

The energies are extracted from the generalized
eigenvalues
\begin{equation}
\lambda_n(t,t_0) = e^{ -E_n(t - t_0) }\, .
\label{eq:geneigen}
\end{equation}
We use the effective mass from Eq.~(\ref{eq:geneigen})
as a check on the masses from the fits to 
Eq.~(\ref{eq:corrmodel}).

\begin{table*}[tb]
\centering
\begin{tabular}{|c|| c|c|| c|c|} \hline
  $\beta$ & \multicolumn{2}{|c|}{light} & 
\multicolumn{2}{|c|}{strange} \\
 & $(\gamma_5 \times \gamma_5)$ & $(\gamma_5 \times 1) $
 & $(\gamma_5 \times \gamma_5)$ & $(\gamma_5 \times 1) $\\
\hline
6.75  & 0.183(1)   & 0.325(2) & 0.387(2)  & 0.4683(8) \\ 
7.095 & 0.1632(7)  & 0.206(1) & 0.3283(7) & 0.3505(6) \\ \hline
\end{tabular}
\caption{Masses for pseudoscalar mesons in 
lattice units from the connected correlators.}
\label{tb:ConnMass}
\end{table*}

We used the recently updated value of 
$r_0$=0.4661(38) fm~\cite{Davies:2009tsa} to determine
the lattice spacing from our measurements of
$r_0/a$~\cite{Richards:2010ck}.
The inverse lattice
spacing determined from $r_0$ is 1.61 and 2.14  $\mbox{GeV}$
for the coarse and fine ensembles respectively.
With our estimates of the lattice spacing we find the value
of the Goldstone pion mass on the coarse and fine ensemble to be
295 and 349 MeV respectively.

The mass of the connected strange-strange pseudoscalar
meson 
$m_{\eta_s}$ = 0.6858(40) GeV, has recently been
determined by the 
HPQCD collaboration~\cite{Davies:2009tsa}. 
By using $m_{\eta_s}^2 \propto m_s$ we find that
the mass of the strange quark is mistuned by -20\%
on the coarse ensemble and +5\% on the fine ensemble.

The masses in table ~\ref{tb:ConnMass} can be used to check
the reduction of taste breaking as the lattice spacing is 
reduced~\cite{Aubin:2004wf}. Defining
\begin{equation}
d  \equiv  (m_{(\gamma_5 \times 1)}^2 - m_{(\gamma_5 \times \gamma_5)}^2
) r_0^2,
\label{eq:tastecheck}
\end{equation}
we find $d_{\rm fine} / d_{\rm coarse} = 0.39(1) $.
The MILC collaboration obtained ~\cite{Aubin:2004wf} 
$d_{\rm fine} / d_{\rm coarse} = 0.38(3)$ at similar parameters.
Note that MILC use $r_1$ in equation~\ref{eq:tastecheck},
but the conversion between $r_1$ and $r_0$ drops out 
in the ratio.

\section{Analysis of disconnected loops}
\label{se:loops}

In~\cite{Gregory:2007ev} we reported on methods to compute disconnected
diagrams with improved staggered fermions. 
We found that
the  technique proposed by Kilcup and 
Venkataraman~\cite{Venkataraman:1997xi}
was the most efficient one of those we tested 
for computing the disconnected diagrams of 
$(\gamma_5 \times 1) $ and $(1 \times 1) $
operators. We therefore used that with 32 Gaussian noise vectors
in this calculation.

The integrated autocorrelation time for the light and strange 
$(\gamma_5 \times 1) $ loops are 42(7) and 24(3) in units
of trajectories respectively. This can be compared
with the autocorrelation times of the connected correlators
for Euclidean time separation 3 of 14(4) and 12(5) for the light and
strange quarks respectively.

\subsection{Ratio of disconnected to connected diagrams.}

%
%

 \begin{figure}
\begin{center}
\includegraphics[scale=0.3,angle=0]{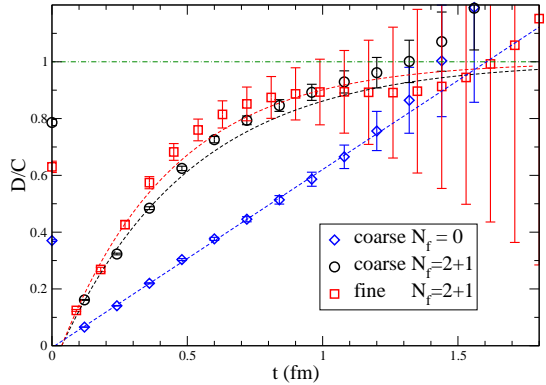}
\end{center}
 \caption {
$D(t)/C(t)$ ratio for the coarse and fine ensembles. We 
also include the result from a 
quenched lattice QCD calculation~\cite{Gregory:2007ev}
with 
a lattice spacing close to the coarse ensemble.
 }
\label{fig:DbyCsummary}
 \end{figure}

An important test of the staggered formalism is to check the 
long time behavior of the $D(t)/C(t)$ 
ratio of the correlators in Eq.~(\ref{eq:coor}). In quenched
QCD, the $D(t)/C(t)$ ratio rises linearly with time but,
in unquenched QCD, the ratio should tend 
to 1 for large times. 
This test was performed for clover 
fermions in \cite{Allton:2004qq}.
In our earlier study ~\cite{Gregory:2007ev}, 
the statistics were not large 
enough to determine the large time behavior of 
the ratio in unquenched QCD.
In Fig.~\ref{fig:DbyCsummary} 
we plot the $D(t)/C(t)$ ratio
for the coarse and fine ensembles respectively,
as well as a quenched ensemble with the same
lattice spacing as the coarse ensemble.

With the higher statistics afforded by the two new ensembles
there is now a clearer difference between the 
quenched and unquenched $D(t)/C(t)$ ratios. 
The $D(t)/C(t)$
ratio for the two unquenched seem to be reaching a
plateau close to 1. As explained in Sec.~\ref{se:simulations},
the staggered ratio depends on the
\lq valence rooting\rq{} factors of 1/4, so this is a good
test of the staggered formalism.

\subsection{Fit results for the masses}

\begin{table*}[tb]
\centering
\begin{tabular}{|c||c|c||c|c|c|c|} \hline
$\beta$ & $a m_{\eta}$ & $a m_{\eta^\prime}$ &
$N_{\rm exp}$ & $t_{\rm min}$-$t_{\rm max}$  & SVD cut & $\chi^2/{\rm dof}$ \\
\hline
6.75  & 0.410(3) & 0.52(1) & 4 & 5-16 & $10^{-5}$ & 1.1\\
7.095 & 0.296(3) & 0.46(2) & 4 & 6-20 & $10^{-5}$ & 0.73\\
\hline
\end{tabular}
\caption{Masses for taste singlet pseudoscalar mesons in 
lattice units. The errors are from the jackknife method. 
The corresponding fit methods are described in the text.} 
\label{tb:DisMass}
\end{table*}

\begin{figure}
\begin{center}
\includegraphics[scale=0.3,angle=0]{PS_FILES/coarse_dt2.new.eps}
\end{center}
 \caption {
Comparison of fit results (horizontal bands) to effective masses 
(with $\Delta t =t - t_0 = 2$) from the variational method for a $4\times 4$ correlator 
matrix on the coarse ensemble.
 }
\label{fig:varyCoarse}
 \end{figure}

\begin{figure}
\begin{center}
\includegraphics[scale=0.3,angle=0]{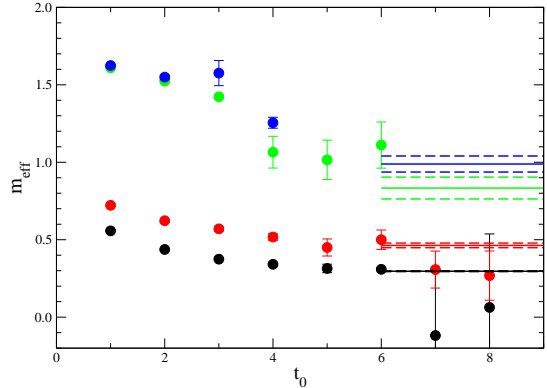}
\end{center}
 \caption {
Comparison of fit results (horizontal bands) to effective masses 
(with $\Delta t=t - t_0  =2$) from the variational method for a $4\times 4$ correlator 
matrix on the fine ensemble.
 }
\label{fig:varyFine}
 \end{figure}

We plot the fit results against the effective masses
from the variational method (\ref{eq:geneigen})
in Figs.~\ref{fig:varyCoarse}
and ~\ref{fig:varyFine}. Here, we set $\Delta = t-t_0 = 2$ in (\ref{eq:geneigen}).
The parameters from the final factorising 
fits are in given Table~\ref{tb:DisMass}.
In arriving at these, we varied  $t_{\rm min}$
and  $N_{\rm exp}$ looking for a sweet spot of stability with respect to
these, for $\chi^2/{\rm dof} \lsim 1$, and for error bars small enough to
distinguish three states. Plateaux in $t_{\rm min}$  were small due to the
competition between excited states at small $t$ and the onset of large
noise/signal at larger $t$.
An SVD cut of around $10^{-5}$ was necessary to get fits with reasonable
confidence levels.

Note that the statistical errors on the final mass parameters deduced
from these correlated fits are generally somewhat less than the
individual variational effective masses calculated at each value of $t$,
even though the same number of configurations were used.
The global factorizing fits include all the data from the different flavor
and fuzzing channels simultaneously, and from a large range of $t$-values. 
They also take correlations into account. 
The individual effective mass estimates from (\ref{eq:geneigen}), on the other 
hand, are each determined by relatively small samples of this data set --- 
uncorrelated measurements from two time-slices yield independent estimates 
for each eigenvalue.

We found it necessary to restrict $t_{\rm max}$ to moderate values 
(16 and 20) since, at larger values, the correlators can fluctuate below
zero. 
We do not attribute this large-$t$ behavior to unphysical effects. 
Rather, the unique statistical properties of disconnected correlators 
mean that a small number of configurations at the tail of the 
distribution can cause large fluctuations of the mean, as discussed in
\cite{Gregory:2007ev}.

We associate the lowest state with $\eta$ and the first excited state
with the $\eta'$.

\section{Comparison with experiment and other lattice calculations}
\label{se:expt}

Given that we only have two masses at two different 
lattice spacings, we cannot make a controlled continuum or
chiral extrapolation. However, we do wish to check
that the resulting parameters are physically
reasonable.

In lattice QCD calculations with two flavors of sea quarks, 
it was found that the mass of the
$\eta_2$ light flavor singlet pseudoscalar meson
had very little mass dependence~\cite{Jansen:2008wv}.

\begin{figure}
\begin{center}
\includegraphics[scale=0.3,angle=0]{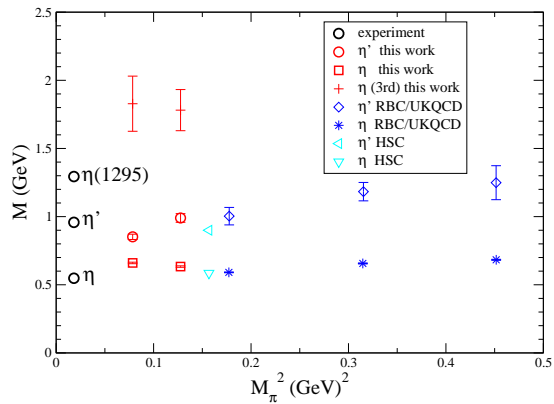}
\end{center}
 \caption {
Summary of mass dependence of the $\eta$ and $\eta^\prime$ as a
function of the square of the pion mass.
We also include data from
the UKQCD/RBC~\cite{Christ:2010dd}, 
Hadron Spectrum Collaboration~\cite{Dudek:2011tt}.
The state listed as $\eta(\hbox{3rd})$ is the next excited state in
our fits. 
}
\label{fig:etasummary}
 \end{figure}

\begin{figure}
\begin{center}
\includegraphics[scale=0.3,angle=0]{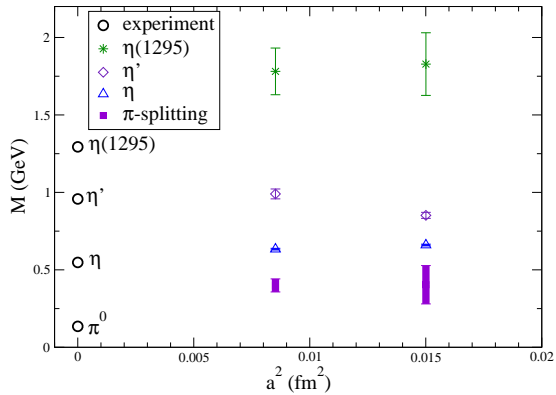}
\end{center}
 \caption {
Summary of the $\eta$ and $\eta^\prime$ as a
function of the square of the lattice spacing. We also
include the mass of the connected $(\gamma_5 \times 1)$
operators to show the scale of the lattice spacing errors.
 }
\label{fig:etasummaryLATT}
 \end{figure}

In Fig.~\ref{fig:etasummary} we plot our results for
the masses of the $\eta$ and $\eta^\prime$ mesons as a 
function of the square of the
pion mass. We also include the recent results from the UKQCD/RBC
collaboration~\cite{Christ:2010dd} and the 
hadron spectrum collaboration~\cite{Dudek:2011tt}.
Note that while the lattice spacing is larger for
the coarse ensemble than the fine ensembles, the 
light quark mass is smaller for the coarse ensemble than the fine.

The mass from the $(\gamma_5 ,1)$ pion operator
is different from the Goldstone  $(\gamma_5 ,\gamma_5)$
pion operator, because of taste breaking. In Fig.~\ref{fig:etasummaryLATT}
we plot, as a function of the square of the lattice spacing,
the masses of the $\eta$ and $\eta^\prime$ mesons along
with the mass splitting between the $(\gamma_5 ,1)$ and
$(\gamma_5 , \gamma_5)$ pions.

\section{$\eta$-$\eta^\prime$ mixing}
\label{se:mixing}


In a quantum mechanics
analysis of the mixing of the 
$\eta$ and $\eta^\prime$~\cite{Isgur:1976qg,Gilman:1987ax}, 
the starting point is
two bare flavor singlet pseudoscalar mesons.
As interactions are introduced, the two
flavor singlet mesons mix and a mixing angle
is introduced. This mixing angle is a parameter
in the various decays of the $\eta$ and $\eta^\prime$
mesons. 

In unquenched lattice QCD calculations it is not
clear how to start from unmixed states.
It is possible to estimate the elements of the
mixing matrix in quenched and 
partially quenched QCD~\cite{McNeile:2000hf,Schilling:2004kg}.
We also note that the properties of the 
connected strange-strange pseudoscalar meson determined
by the HPQCD collaboration~\cite{Davies:2009tsa}
are also useful for developing a partially quenched
mixing analysis~\cite{McNeile:2000hf}.

In QCD, the mixing between the 
$\eta$ and $\eta^\prime$ mesons is understood
at the amplitude level. In phenomenological
analysis of decays, the mixing is parametrized in 
terms of decay constants and 1 or 2 angles.
Many of the analyses of the decays also include
additional strong interaction physics,
such as form factors~\cite{Thomas:2007uy}. 

Di Donato et al.~\cite{DiDonato:2011kr}
and~\cite{Fleischer:2011ib} Fleischer et al.
have recently discussed the importance
of the mixing of $\eta$ and  $\eta^\prime$ mesons on various
decays.
Shore and collaborators ~\cite{Shore:2007yn}
have derived equations
relating the masses and decay constants of the $\eta$ and $\eta^\prime$
to the topological susceptibility.
%
%

The phenomenological analysis of $\eta$-$\eta^\prime$ 
mixing is usually described
in terms of the axial decay constants~\cite{Feldmann:1999uf}. 
To study the
decay constants of flavor singlet meson operators
with staggered fermions requires the $(\gamma_4 \gamma_5 , 1)$
pion operators. These were not included in our calculation.
Since there is no Ward identity for the $(\gamma_4 \gamma_5 , 1)$
operators, renormalization factors would need to be 
computed~\cite{Aoki:1999av}.
The connection between $\eta-\eta^\prime$ mixing with axial currents
or pseudoscalar currents is discussed in
section 3.8 of Feldmann's review~\cite{Feldmann:1999uf}.
To first order the mixing angle is the same with axial
or pseudoscalar currents.

There are two popular bases in which to express $\eta$/$\eta'$ states.
One basis is the 
$SU(3)$ basis:
\begin{eqnarray}
| \eta_0 \rangle & = & \frac{1}{\sqrt{3}}
\left(
| u\overline{u} \rangle + | d\overline{d} \rangle + | s\overline{s} \rangle
\right)
\nonumber \\
| \eta_8 \rangle & = & \frac{1}{\sqrt{6}}
\left(
| u\overline{u} \rangle + | d\overline{d} \rangle -
2 |s\overline{s}\rangle
\right)\, .
\end{eqnarray}
The $\eta$ and $\eta^\prime$ mesons states are then linear
combinations of the basis states:
\begin{equation}
\begin{pmatrix}
\mid \eta \rangle\\
\mid \eta' \rangle
\end{pmatrix}
=
\begin{pmatrix}
\cos\theta_P & -\sin\theta_P\\
\sin\theta_P & \cos\theta_P\\
\end{pmatrix}
\begin{pmatrix}
\mid \eta_8 \rangle\\
\mid \eta_0 \rangle
\end{pmatrix}
\,.
\end{equation}
The $\eta$ and $\eta^\prime$ mesons can also 
be described using the quark flavor basis:
\begin{eqnarray}
| \eta_q \rangle & = & \frac{1}{\sqrt{2}}\left( | u\overline{u}
\rangle + |d\overline{d} \rangle \right)
\nonumber \\
| \eta_s \rangle & = & |s\overline{s}\rangle
\end{eqnarray}
\begin{equation}
\begin{pmatrix}
| \eta \rangle\\
| \eta' \rangle
\end{pmatrix}
=
\begin{pmatrix}
\cos\phi_P & -\sin\phi_P\\
\sin\phi_P & \cos\phi_P\\
\end{pmatrix}
\begin{pmatrix}
| \eta_q \rangle\\
| \eta_s \rangle
\end{pmatrix}\, .
\end{equation}

The two mixing angles are related
to each other:
\begin{equation}
\theta_P = \phi_P -\arctan\sqrt{2} = \phi_P-54.7^{\circ}\, .
\label{eq:basisconv}
\end{equation}

In lattice QCD calculations it is not possible to start
from pure basis states such as $| \eta_q \rangle $
and $| \eta_s  \rangle$. Instead the mixing is described
in terms of amplitudes. For example we can define
an amplitude $a_{\eta_8 \eta}$ via
\begin{equation}
a_{\eta_8 \eta} = \langle 0 \mid \eta_8 \mid \eta \rangle.
\end{equation}
Other amplitudes, for different interpolating operators
and mesons, are similarly defined. 
For example, the amplitudes $a_{\eta_q \eta}$ and
$a_{\eta_s \eta}$
are parameters in the fit model in Eq.~(\ref{eq:corrmodel}),
with numerical indices replaced by the names of the meson
or interpolating operator.

With two possible interpolating operators and two
states, there are in principle four possible decay constants.
The discussion of mixing in quantum mechanics
suggests that these four amplitudes can be parameterized
in terms of two decay constants and a mixing angle.

It has been found that, with
$SU(3)$-breaking, there cannot be one single mixing angle
in the $SU(3)$ basis~\cite{Feldmann:1998vh}. For
example from a review of the literature, 
Feldmann~\cite{Feldmann:1999uf} notes a spread
of 10 degrees in the mixing angles extracted from
experiment. In the $SU(3)$ flavor basis the mixing of decay constants
is expressed by
\begin{equation}
\begin{pmatrix}
a_{8\eta}  & a_{0\eta}\\
a_{8\eta'} & a_{0\eta'}
\end{pmatrix}
=
\begin{pmatrix}
f_8\cos\theta_8 & -f_0\sin\theta_0 \\
f_8\sin\theta_8 & f_0\cos\theta_0 \\
\end{pmatrix}\, .
\end{equation}
The decay constants can also be computed in
the quark basis:
\begin{equation}
\begin{pmatrix}
a_{q\eta} & a_{s\eta} \\
a_{q\eta'} & a_{s\eta'}
\end{pmatrix}
=
\begin{pmatrix}
f_q\cos\phi_q & -f_s\sin\phi_s \\
f_q\sin\phi_q & f_s\cos\phi_s \\
\end{pmatrix}\, .
\label{eq:quarkMixing}
\end{equation}

To extract the mixing angles from the matrix
in Eq.~(\ref{eq:quarkMixing}), the following
combinations can be used:
\begin{equation}
\tan \phi_q^{\rm est} = \frac{a_{q\eta'}}{a_{q\eta} }
\label{eq:phiq}
\end{equation}
\begin{equation}
\tan \phi_s^{\rm est} = -\frac{a_{s\eta} }{a_{s\eta'} }\, . 
\label{eq:phis}
\end{equation}

There are arguments~\cite{Leutwyler:1997yr,Feldmann:1999uf}
that suggest $\phi_q \approx \phi_s$. 
Feldmann~\cite{Feldmann:1999uf}
reviews
the determination of the mixing angles from
various processes and he finds 
$\mid \phi_q - \phi_s \mid < 5$ degrees.
The UKQCD/RBC colaborations use perturbation theory to
identify conditions where there is 
only one angle~\cite{Christ:2010dd}.

\begin{figure}
\begin{center}
\includegraphics[scale=0.3,angle=0]{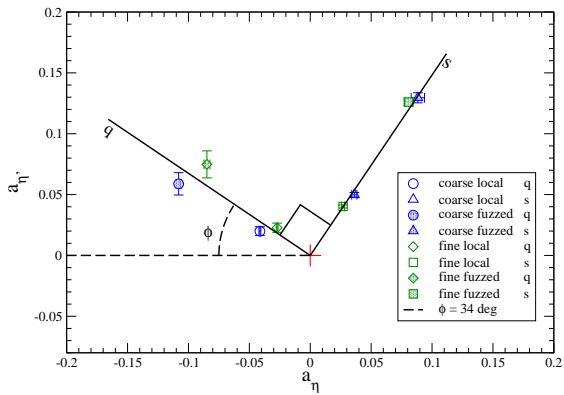}
\end{center}
\caption{The fit amplitudes plotted in couplets, $(a_\eta,a_{\eta'})$,
with the $\eta$ amplitude as 
the horizontal coordinate and the $\eta'$ amplitude as the vertical component, 
is a graphical representation of Eq.~(\ref{eq:quarkMixing}). 
This illustrates 
several nontrivial results, including the consistency of the mixing angles 
from local (open) and fuzzed (filled) sources, and the consistency of mixing 
angles from coarse and fine simulations. The extent to which the light quark and 
strange quark branches lie at right angles to each other is a measure of the 
aptness of the single mixing angle description.
We use lattice units; the inclusion of renormalization factors will not 
change the angular distribution.
}
\label{fg:mix_angle}
\end{figure}

In Fig.~\ref{fg:mix_angle} we show the fit amplitudes plotted in couplets
$(a_\eta,a_{\eta'})$ demonstrating the mixing angles in the quark flavor 
basis as in Eq.~(\ref{eq:quarkMixing}).

Because of SU(3) symmetry breaking, $f_q$ is not equal to
$f_s$ in Eq.~(\ref{eq:quarkMixing}), hence a rotation
of the interpolating basis from the quark to the 
flavor basis will no longer give a matrix of
the same form as Eq.~(\ref{eq:quarkMixing}).
The use of mixing angles to parametrize the mixing matrix 
is similar to the
polar decomposition of a 
matrix~\cite{Higham:2010zz}, where a matrix can be 
written as the product of a Hermitian matrix and a
unitary matrix. In this application the Hermitian matrix 
is actually diagonal.

The RBC/UKQCD 
and Hadron Spectrum Collaboration extract a
single angle by computing the following angle
\begin{equation}
\tan^2 \phi^{\rm est} = - \frac{a_{q\eta'} a_{s\eta} }{a_{q\eta}
  a_{s\eta'} }\, ,
\label{eq:phiest}
\end{equation}
so that 
$\tan \phi^{\rm est}$ is the geometric mean of 
$\tan \phi_q^{\rm est}$ and $\tan \phi_s^{\rm est}$.

\begin{table*}[tb]
\centering
\begin{tabular}{|c||c|c|c||c|c|} \hline
$\beta$ & $\phi_q^{\rm est}$ & $\phi_s^{\rm est}$ & $\phi^{\rm est}$ & 
$\phi^{\rm fit}$ & $\chi^2/dof$ \\
\hline
6.75  &  25(4) & 36(2) & 31(4)  & 34(3) & $8.2/3$ \\
7.095 &  40(5) & 34(2) &  37(3) & 34(2)  & $3.7/3$ \\
\hline
\end{tabular}
\caption{$\eta$-$\eta^\prime$ mixing angles in degrees
defined and determined as described in the text. 
}
\label{tb:Mixetaetaprime}
\end{table*}

The relation $\phi_q \approx \phi_s$
is an assumption, under which the four possible
parameters are reduced to three, hence it is interesting
to test it using lattice QCD. 
We have determined the mixing angles 
using different methods and present the results
in Table~\ref{tb:Mixetaetaprime}. 
The angles $\phi_q^{\rm est}$, $\phi_s^{\rm est}$ and $\phi^{\rm est}$     
were obtained from the fits to Eq.~(\ref{eq:corrmodel}),
using Eqs.(\ref{eq:phiq}), (\ref{eq:phis}) 
and~(\ref{eq:phiest}). Local operators were used for this.
We also found similar results using fuzzed operators.

The angle $\phi^{\rm fit}$, 
in Table~\ref{tb:Mixetaetaprime},
was determined by simultaneously fitting the 
smeared and local matrix of amplitudes
in Eq.~(\ref{eq:quarkMixing}) using a single angle
$\phi^{\rm fit}=\phi_q=\phi_s$, with two decay
constants for the local matrix, and
two decay constants for the fuzzed
correlators.

One cannot confidently conclude from the
results in Table~\ref{tb:Mixetaetaprime},
that a single mixing angle does not
describe the data.
As an exercise we ignored
the statistical errors on the amplitudes
from the $\beta=7.095$ ensemble and rotated the matrix
in Eq.~\ref{eq:quarkMixing}
by an angle $\alpha$ in the interpolating
operator basis and computed $\phi_q^{\rm est}$
and  $\phi_s^{\rm est}$. Around 9 degrees from
the quark basis we found $\phi_q^{\rm est} = \phi_s^{\rm est}$
as a function of $\alpha$.
This shows as expected that the quark basis is
a good basis for having a single mixing angle.
This method can be used with future more precise lattice
QCD calculations, to determine the optimal basis
of interpolating operators, such that the amplitudes
can be described by a single mixing angle.

\begin{figure}
\begin{center}
\includegraphics[scale=0.3,angle=0]{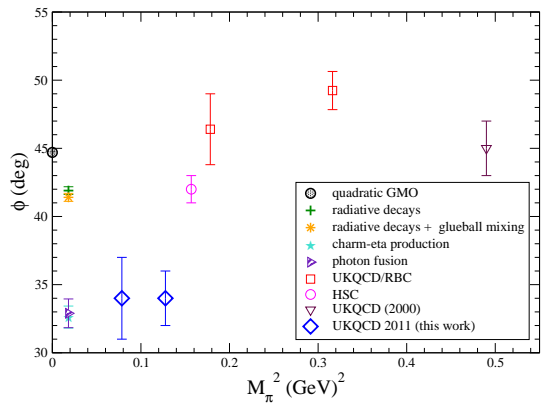}
\end{center}
\caption{A summary of $\eta$-$\eta^\prime$ mixing angles 
as a function of the square of pion mass. We also include data from
the UKQCD/RBC~\cite{Christ:2010dd}, 
Hadron Spectrum Collaboration~\cite{Dudek:2011tt}, 
and UKQCD collaborations~\cite{McNeile:2000hf}.
The UKQCD/RBC number is quoted in the  SU(3) basis, 
so we have used equation~\ref{eq:basisconv}  
to convert it to the quark basis.
}
\label{fg:thetasummary}
\end{figure}

In Fig.~\ref{fg:thetasummary} we plot the mixing 
angle in the quark basis ($\phi^{\rm fit}$),
obtained using the fit method,
as a function of the square of the pion mass.
We also include some experimental numbers for the 
mixing angle from the summary in~\cite{DiDonato:2011kr}.
We see that our results from staggered fermions are
qualitatively in agreement with the results of the
other two lattice groups and experiment.

\section{Conclusions}
\label{se:conclusions}

We have reported a lattice QCD calculation
of the masses and mixing of the $\eta$-$\eta^\prime$ mesons.
Our key results for the 
masses (Fig.~\ref{fig:etasummaryLATT})
and mixing angles (Fig.~\ref{fg:thetasummary}),
and the ratio plot (Fig.~\ref{fig:DbyCsummary} )
do not give any grounds for concern over the 
validity of the staggered fermion formulation.
Our results are qualitatively in agreement with experiment
and with lattice analyses using other formulations.

Using a technique specific to staggered fermions~\cite{Venkataraman:1997xi}
we were able to compute the disconnected diagrams at a
cost roughly 10 times the cost of the connected
correlators. However, as always, higher statistics are required for
lattice QCD calculations which require disconnected diagrams.

The main shortcoming of the calculation is that we have used 
only one quark mass at each of two different lattice spacings.
It is clearly important to extrapolate
the masses of the $\eta$  and $\eta^\prime$
mesons to the physical quark masses
and to take the continuum limit.
To do this, it may be better to use a staggered fermion formalism
with reduced flavor symmetry breaking. The 
MILC collaboration~\cite{Bazavov:2010ru}
and HOT collaboration~\cite{Bazavov:2011nk}
have started to simulate 
the HISQ~\cite{Follana:2006rc}
fermion action, which uses an additional
level of fat links over the ASQTAD action so reducing
the mass splitting between the masses of the 
$(\gamma_5 \times 1)$ and
$(\gamma_5 \times \gamma_5)$ pions.

The stout link staggered action used 
by Aoki et al.~\cite{Aoki:2005vt,Aoki:2009sc}
could also be used, since this formalism has reduced
taste splitting compared to ASQTAD and the quark masses
are within 10\% of the physical light and strange
masses.

For the phenomenology of $\eta$ and $\eta^\prime$
it is important to study the decay constants
and mixing angle of the $\eta$ and $\eta^\prime$ mesons.
Although staggered fermions are computationally
cheap and have a Ward identity for the Goldstone pion
operator,
it is not clear whether we can obtain 
a strong enough signal for the $(\gamma_4 \gamma_5 \times 1)$
operator to be able to 
extract decay constants~\cite{Aoki:1999av}.


Even more challenging would be to study the mixing
of the the $\eta^\prime$ and $\eta$ with the
pseudoscalar glueball. This requires an extension of the 
formalism for mixing described in section~\ref{se:mixing}.
In principle, the mixing angle between the $\eta$ and
$\eta^\prime$ and the pseudoscalar glueball can be 
obtained by fitting various branching ratios. 
However,
a clear picture from the phenomenological approach
has yet to 
emerge~\cite{DiDonato:2011kr,Cheng:2008ss,Shore:2006mm,Fleischer:2011ib}, 
so a first
principle calculation from lattice QCD would be 
valuable. 
The KLOE experiment~\cite{AmelinoCamelia:2010me} has
analysed their branching fraction data for vector meson
radiative decay to pseudoscalar mesons with a model
and obtained an estimate of $\eta^\prime$-glueball mixing 
which is three sigma from zero.
The KLOE-2 experiment plans to reduce the experimental errors
on the analysis~\cite{AmelinoCamelia:2010me}.
Given that we have already 
published~\cite{Richards:2010ck} a 
study of the pseudo-scalar glueball on these ensembles,
we could at the very least include the 
gluonic and fermionic pseudoscalar interpolating operators
in a single variational calculation (as was done for scalar
operators in~\cite{Hart:2006ps}). 

This calculation would be interesting from
a nonperturbative QCD perspective, but also would
be important for exploring CP violation in the decays of
$B$ and $B_s$ mesons to final states that include the
$\eta$ and $\eta^\prime$.

Lattice QCD calculations of the $\eta$ and $\eta^\prime$ mesons
involve an interesting mixture of: toplogy, anamolies,
mixing via quark loops, and have important applications
to phenomenology.
The relatively high cost of lattice QCD calculations of the
$\eta$ and $\eta^\prime$ mesons 
has meant that they
are not as highly developed
as the calculation of flavor 
nonsinglet hadrons~\cite{Bazavov:2009bb}.
Since the computational costs are becoming less exorbitant, 
we expect that lattice calculations of 
the $\eta$ and $\eta^\prime$ mesons may soon become
precision studies.

\section{Acknowledgments}

We thank Steven Miller, Zbyszek Sroczynski
for participating in the early part of this project.
We thank Chulwoo Jung and Mike Clark for providing optimized code used in
this project. We also thank Alistair Hart, Andreas Kronfeld,
Konstantin Ottnad and Carsten Urbach for their helpful comments.
We used the North West grid, Scotgrid and Bluegene/P
at Daresbury.
The measurements used the Chroma software system~\cite{Edwards:2004sx}.
The configurations were generated on the 
QCDOC~\cite{Boyle:2005gf}. EBG was supported at UCY by
Cyprus Research Promotion Foundation grant 
$\Delta IE\Theta NH\Sigma/\Sigma TOXO\Sigma/0308/07$.

\bibliographystyle{h-physrev5}
\bibliography{longrun}

\begin{thebibliography}{10}

\bibitem{Feldmann:1999uf}
T.~Feldmann,
\newblock Int. J. Mod. Phys. {\bf A15}, 159 (2000), hep-ph/9907491.

\bibitem{Shore:2007yn}
G.~M. Shore,
\newblock Lect. Notes Phys. {\bf 737}, 235 (2008), arXiv:hep-ph/0701171.

\bibitem{Kawarabayashi:1980dp}
K.~Kawarabayashi and N.~Ohta,
\newblock Nucl.Phys. {\bf B175}, 477 (1980).

\bibitem{DiDonato:2011kr}
C.~Di~Donato, G.~Ricciardi, and I.~Bigi,
\newblock (2011), arXiv:1105.3557.

\bibitem{Fleischer:2011ib}
R.~Fleischer, R.~Knegjens, and G.~Ricciardi,
\newblock (2011), arXiv:1110.5490.

\bibitem{Moskal:2011ve}
P.~Moskal,
\newblock (2011), arXiv:1102.5548.

\bibitem{Adam:2004ch}
WASA-at-COSY Collaboration, H.-H. Adam {\em et~al.},
\newblock (2004), arXiv:nucl-ex/0411038.

\bibitem{AmelinoCamelia:2010me}
G.~Amelino-Camelia {\em et~al.},
\newblock Eur.Phys.J. {\bf C68}, 619 (2010), arXiv:1003.3868.

\bibitem{Unverzagt:2009vm}
Crystal Ball at MAMI Collaboration, M.~Unverzagt,
\newblock Nucl.Phys.Proc.Suppl. {\bf 198}, 174 (2010), arXiv:0910.1331.

\bibitem{Orginos:1998ue}
MILC, K.~Orginos and D.~Toussaint,
\newblock Phys. Rev. {\bf D59}, 014501 (1999), hep-lat/9805009.

\bibitem{Orginos:1999cr}
MILC, K.~Orginos, D.~Toussaint, and R.~L. Sugar,
\newblock Phys. Rev. {\bf D60}, 054503 (1999), hep-lat/9903032.

\bibitem{Davies:2003ik}
HPQCD, C.~T.~H. Davies {\em et~al.},
\newblock Phys. Rev. Lett. {\bf 92}, 022001 (2004), hep-lat/0304004.

\bibitem{Gregory:2009hq}
E.~B. Gregory {\em et~al.},
\newblock Phys. Rev. Lett. {\bf 104}, 022001 (2010), arXiv:0909.4462.

\bibitem{Bazavov:2009bb}
A.~Bazavov {\em et~al.},
\newblock Rev.Mod.Phys. {\bf 82}, 1349 (2010), arXiv:0903.3598.

\bibitem{Creutz:2007yg}
M.~Creutz,
\newblock Phys.Lett. {\bf B649}, 230 (2007), arXiv:hep-lat/0701018.

\bibitem{Creutz:2008nk}
M.~Creutz,
\newblock PoS {\bf CONFINEMENT8}, 016 (2008), arXiv:0810.4526.

\bibitem{Durr:2005ax}
S.~Durr,
\newblock PoS {\bf LAT2005}, 021 (2006), arXiv:hep-lat/0509026.

\bibitem{Sharpe:2006re}
S.~R. Sharpe,
\newblock PoS. {\bf LAT2006}, 022 (2006), hep-lat/0610094.

\bibitem{Creutz:2007rk}
M.~Creutz,
\newblock PoS {\bf LAT2007}, 007 (2007), arXiv:0708.1295.

\bibitem{Kronfeld:2007ek}
A.~S. Kronfeld,
\newblock PoS {\bf LAT2007}, 016 (2007), arXiv:0711.0699.

\bibitem{Follana:2004sz}
HPQCD, E.~Follana, A.~Hart, and C.~T.~H. Davies,
\newblock Phys. Rev. Lett. {\bf 93}, 241601 (2004), hep-lat/0406010.

\bibitem{Follana:2005km}
HPQCD Collaboration, UKQCD Collaboration, E.~Follana, A.~Hart, C.~Davies, and
  Q.~Mason,
\newblock Phys.Rev. {\bf D72}, 054501 (2005), arXiv:hep-lat/0507011.

\bibitem{Donald:2011if}
G.~C. Donald, C.~T.~H. Davies, E.~Follana, and A.~S. Kronfeld,
\newblock (2011), arXiv:1106.2412.

\bibitem{Durr:2004as}
S.~Durr, C.~Hoelbling, and U.~Wenger,
\newblock Phys. Rev. {\bf D70}, 094502 (2004), hep-lat/0406027.

\bibitem{Bazavov:2010xr}
MILC, A.~Bazavov {\em et~al.},
\newblock Phys. Rev. {\bf D81}, 114501 (2010), arXiv:1003.5695.

\bibitem{Aubin:2004wf}
C.~Aubin {\em et~al.},
\newblock Phys. Rev. {\bf D70}, 094505 (2004), hep-lat/0402030.

\bibitem{Bernard:2007qf}
C.~Bernard, C.~E. DeTar, Z.~Fu, and S.~Prelovsek,
\newblock Phys. Rev. {\bf D76}, 094504 (2007), arXiv:0707.2402.

\bibitem{Bernard:2003jd}
C.~Bernard {\em et~al.},
\newblock Phys. Rev. {\bf D68}, 074505 (2003), arXiv:hep-lat/0301024.

\bibitem{Gregory:2007ev}
E.~B. Gregory, A.~C. Irving, C.~M. Richards, and C.~McNeile,
\newblock Phys. Rev. {\bf D77}, 065019 (2008), arXiv:0709.4224.

\bibitem{Richards:2010ck}
UKQCD, C.~M. Richards, A.~C. Irving, E.~B. Gregory, and C.~McNeile,
\newblock Phys. Rev. {\bf D82}, 034501 (2010), arXiv:1005.2473.

\bibitem{Aoki:2006xk}
JLQCD, S.~Aoki {\em et~al.},
\newblock (2006), hep-lat/0610021.

\bibitem{Christ:2010dd}
N.~H. Christ {\em et~al.},
\newblock Phys. Rev. Lett. {\bf 105}, 241601 (2010), arXiv:1002.2999.

\bibitem{Dudek:2011tt}
J.~J. Dudek {\em et~al.},
\newblock Phys. Rev. {\bf D83}, 111502 (2011), arXiv:1102.4299.

\bibitem{Kaneko:2009za}
TWQCD collaboration, JLQCD Collaboration, T.~Kaneko {\em et~al.},
\newblock PoS {\bf LAT2009}, 107 (2009), arXiv:0910.4648.

\bibitem{Bali:2011yx}
QCDSF Collaboration, G.~Bali {\em et~al.},
\newblock (2011), arXiv:1111.4053.

\bibitem{Ottnad:2011mp}
K.~Ottnad, C.~Urbach, C.~Michael, and S.~Reker,
\newblock (2011), arXiv:1111.3596.

\bibitem{Jansen:2008wv}
ETM, K.~Jansen, C.~Michael, and C.~Urbach,
\newblock Eur. Phys. J. {\bf C58}, 261 (2008), arXiv:0804.3871.

\bibitem{Chen:2000bu}
D.~Chen {\em et~al.},
\newblock Nucl. Phys. Proc. Suppl. {\bf 94}, 825 (2001), hep-lat/0011004.

\bibitem{Clark:2006fx}
M.~A. Clark and A.~D. Kennedy,
\newblock Phys. Rev. Lett. {\bf 98}, 051601 (2007), arXiv:hep-lat/0608015.

\bibitem{Clark:2006wp}
M.~A. Clark and A.~D. Kennedy,
\newblock Phys. Rev. {\bf D75}, 011502 (2007), arXiv:hep-lat/0610047.

\bibitem{Richards:2009thesis}
C.~Richards,
\newblock {\em A high statistics study of the scalar singlet states in lattice
  QCD},
\newblock PhD thesis, University of Liverpool, 2009.

\bibitem{Gregory:2005yr}
E.~B. Gregory, A.~C. Irving, C.~McNeile, S.~Miller, and Z.~Sroczynski,
\newblock PoS {\bf LAT2005}, 027 (2006), hep-lat/0510066.

\bibitem{Beane:2009ky}
S.~R. Beane {\em et~al.},
\newblock (2009), arXiv:0903.2990.

\bibitem{Wilcox:1999ab}
W.~Wilcox,
\newblock (1999), hep-lat/9911013.

\bibitem{Alexandrou:2011ar}
C.~Alexandrou, K.~Hadjiyiannakou, G.~Koutsou, A.~O. Cais, and A.~Strelchenko,
\newblock (2011), arXiv:1108.2473.

\bibitem{Endres:2011jm}
M.~G. Endres, D.~B. Kaplan, J.-W. Lee, and A.~N. Nicholson,
\newblock (2011), arXiv:1106.0073.

\bibitem{Endres:2011mm}
M.~G. Endres, D.~B. Kaplan, J.-W. Lee, and A.~N. Nicholson,
\newblock (2011), arXiv:1112.4023.

\bibitem{Bernard:2007ps}
C.~Bernard {\em et~al.},
\newblock PoS {\bf LAT2007}, 090 (2007), arXiv:0710.1118.

\bibitem{Toussaint:2009pz}
MILC, D.~Toussaint and W.~Freeman,
\newblock Phys. Rev. Lett. {\bf 103}, 122002 (2009), arXiv:0905.2432.

\bibitem{KlubergStern:1983dg}
H.~Kluberg-Stern, A.~Morel, O.~Napoly, and B.~Petersson,
\newblock Nucl. Phys. {\bf B220}, 447 (1983).

\bibitem{Lacock:1994qx}
UKQCD, P.~Lacock, A.~McKerrell, C.~Michael, I.~M. Stopher, and P.~W.
  Stephenson,
\newblock Phys. Rev. {\bf D51}, 6403 (1995), arXiv:hep-lat/9412079.

\bibitem{Lepage:2001ym}
G.~P. Lepage {\em et~al.},
\newblock Nucl. Phys. Proc. Suppl. {\bf 106}, 12 (2002), arXiv:hep-lat/0110175.

\bibitem{Michael:1985ne}
C.~Michael,
\newblock Nucl.Phys. {\bf B259}, 58 (1985).

\bibitem{Luscher:1990ck}
M.~Luscher and U.~Wolff,
\newblock Nucl.Phys. {\bf B339}, 222 (1990).

\bibitem{Blossier:2009kd}
B.~Blossier, M.~Della~Morte, G.~von Hippel, T.~Mendes, and R.~Sommer,
\newblock JHEP {\bf 0904}, 094 (2009), arXiv:0902.1265.

\bibitem{Davies:2009tsa}
HPQCD, C.~T.~H. Davies, E.~Follana, I.~D. Kendall, G.~P. Lepage, and
  C.~McNeile,
\newblock (2009), arXiv:0910.1229.

\bibitem{Venkataraman:1997xi}
L.~Venkataraman and G.~Kilcup,
\newblock (1997), hep-lat/9711006.

\bibitem{Allton:2004qq}
UKQCD, C.~R. Allton {\em et~al.},
\newblock Phys. Rev. {\bf D70}, 014501 (2004), hep-lat/0403007.

\bibitem{Isgur:1976qg}
N.~Isgur,
\newblock Phys. Rev. {\bf D13}, 122 (1976).

\bibitem{Gilman:1987ax}
F.~J. Gilman and R.~Kauffman,
\newblock Phys. Rev. {\bf D36}, 2761 (1987).

\bibitem{McNeile:2000hf}
UKQCD, C.~McNeile and C.~Michael,
\newblock Phys. Lett. {\bf B491}, 123 (2000), hep-lat/0006020.

\bibitem{Schilling:2004kg}
K.~Schilling, H.~Neff, and T.~Lippert,
\newblock Lect. Notes Phys. {\bf 663}, 147 (2005), hep-lat/0401005.

\bibitem{Thomas:2007uy}
C.~E. Thomas,
\newblock JHEP {\bf 10}, 026 (2007), arXiv:0705.1500.

\bibitem{Aoki:1999av}
JLQCD, S.~Aoki {\em et~al.},
\newblock Phys. Rev. {\bf D62}, 094501 (2000), arXiv:hep-lat/9912007.

\bibitem{Feldmann:1998vh}
T.~Feldmann, P.~Kroll, and B.~Stech,
\newblock Phys.Rev. {\bf D58}, 114006 (1998), arXiv:hep-ph/9802409.

\bibitem{Leutwyler:1997yr}
H.~Leutwyler,
\newblock Nucl. Phys. Proc. Suppl. {\bf 64}, 223 (1998), arXiv:hep-ph/9709408.

\bibitem{Higham:2010zz}
N.~J. Higham, C.~Mehl, and F.~Tisseur,
\newblock SIAM J.Matrix Anal.Appl. {\bf 31}, 2163 (2010).

\bibitem{Bazavov:2010ru}
MILC, A.~Bazavov {\em et~al.},
\newblock Phys. Rev. {\bf D82}, 074501 (2010), arXiv:1004.0342.

\bibitem{Bazavov:2011nk}
A.~Bazavov {\em et~al.},
\newblock (2011), arXiv:1111.1710.

\bibitem{Follana:2006rc}
HPQCD Collaboration, UKQCD Collaboration, E.~Follana {\em et~al.},
\newblock Phys.Rev. {\bf D75}, 054502 (2007), arXiv:hep-lat/0610092.

\bibitem{Aoki:2005vt}
Y.~Aoki, Z.~Fodor, S.~D. Katz, and K.~K. Szabo,
\newblock JHEP {\bf 01}, 089 (2006), arXiv:hep-lat/0510084.

\bibitem{Aoki:2009sc}
Y.~Aoki {\em et~al.},
\newblock JHEP {\bf 06}, 088 (2009), arXiv:0903.4155.

\bibitem{Cheng:2008ss}
H.-Y. Cheng, H.-n. Li, and K.-F. Liu,
\newblock Phys. Rev. {\bf D79}, 014024 (2009), arXiv:0811.2577.

\bibitem{Shore:2006mm}
G.~M. Shore,
\newblock Nucl. Phys. {\bf B744}, 34 (2006), hep-ph/0601051.

\bibitem{Hart:2006ps}
UKQCD, A.~Hart, C.~McNeile, C.~Michael, and J.~Pickavance,
\newblock Phys. Rev. {\bf D74}, 114504 (2006), hep-lat/0608026.

\bibitem{Edwards:2004sx}
SciDAC, R.~G. Edwards and B.~Joo,
\newblock Nucl. Phys. Proc. Suppl. {\bf 140}, 832 (2005), hep-lat/0409003.

\bibitem{Boyle:2005gf}
P.~Boyle {\em et~al.},
\newblock Nucl.Phys.Proc.Suppl. {\bf 140}, 169 (2005).

\end{thebibliography}

\end{document}